\documentclass[aps,pra,twocolumn,showpacs,floatfix]{revtex4}
\usepackage{epsfig}
\usepackage{graphicx}
\usepackage{dcolumn}
\usepackage{longtable}
\usepackage{amsthm,amsmath}

\begin{document}

\preprint{\today}

\title{Dipole polarizabilities of the transition and post-transition metallic systems}

\author{Yashpal Singh\footnote{yashpal@prl.res.in} and B. K. Sahoo\footnote{bijaya@prl.res.in}}
\affiliation{Theoretical Physics Division, Physical Research Laboratory,
Navrangpura, Ahmedabad - 380009, India}

\begin{abstract}

We investigate the role of the electron correlation effects in the calculations of the electric 
dipole polarizabilities $(\alpha)$ of elements belonging to three different groups of 
periodic table. To understand the propagation of the electron correlation effects 
at different levels of approximations, we employ the relativistic many-body methods developed, based on 
the first principles, at mean-field Dirac-Fock (DF), third order many-body perturbation 
theory (MBPT(3)), random-phase approximation (RPA) and the singly and doubly approximated 
coupled-cluster methods at the linearized (LCCSD) and non-linearized (CCSD) levels. We observe 
variance in the trends of the contributions from the correlation effects in a particular
group of elements through a employed many-body method; however they resemble similar tendency among
the isoelectronic systems. Our CCSD results are within sub-one percent agreement
with the experimental values which are further ameliorated by including the contributions 
from the important triple excitations (CCSD$_p$T method). 
\end{abstract}


\maketitle

\section{Introduction}
Static electric dipole polarizability $(\alpha)$ of an atomic system is the measure of distortion
of the electron cloud when the system is subjected to a stray electric field. Some of the notable
applications with the accurate knowledge of $\alpha$ are in the studies of new generation frequency standards, atomic interactions
in optical lattices, quantum information along with many others in the areas of atomic and
molecular physics \cite{diddams,gill1,gill2,rosenband1,rosenband2,bonin,pethick,madej}. 
Various sophisticated experimental techniques have been exercised to measure $\alpha$ in different
atomic systems having their own merits and disadvantages \cite{hall,molof,miller,schwartz,cronin,ekstrom,amini}. 
Despite of the technological advancements, it is still remained to attain high-precision
measurements of $\alpha$ in the ground states of many atomic systems. In fact, there are also 
some systems where no experimental results are yet available. Nevertheless, accurate
evaluation of $\alpha$ can serve as a good test of the potential of any developed many-body method and
to peruse the underlying interplay of the electron correlation effects in their determination.

A seminal work on the calculations of polarizabilities of the many-electron systems in the {\it ab initio}
framework was first introduced by Dalgarno and his collaborators about more than five decades ago \cite{dalgarno, dalgarno2}. 
Since then variants of advanced many-body methods have been developed and applied 
successfully in the same philosophical stratagem to evaluate this atomic property meticulously.
Examples of few well-known many-body 
methods that are often employed in the studies of $\alpha$ are the random phase approximation (RPA), 
coupled-cluster method in the linear response theory (CCLRT), configuration-interaction (CI) method etc.
\cite{monkhorst,daalgard, kundu,koch,kobayashi,datta,kowalski,hammond}; however many of these methods
are developed in the non-relativistic mechanics. Lim and coworkers have demonstrated, by employing
the coupled-cluster (CC) methods developed using the Cartesian coordinates for the molecular calculations,  
that the relativistic contributions to determine $\alpha$ values are significant, especially in the heavier
atomic systems \cite{lim1,lim2}. In their CC method, the relativistic effects are accounted by using 
a two-component Douglas-Kroll Hamiltonian.

 To encompass both the correlation and the relativistic effects in the $\alpha$ determination of the
closed-shell atomic systems, we have developed a CC method considering the Dirac-Coulomb (DC) Hamiltonian described
by the four-component atomic wave functions in the spherical coordinate system \cite{bijaya2,bijaya1}. 
Ground state $\alpha$ of a number of closed-shell systems have been successfully evaluated using such methodology
in the last couple of years \cite{yashpal-polz, sidhu1, sidhu2}. Moreover, we have set-up methods in the 
third order many-body perturbation theory [MBPT(3)] and RPA in the relativistic formalism with the intention
of including the correlation effects through the first principle calculations as has been employed in \cite{dalgarno,dalgarno2}.
The focus of the present work is to apprehend the role of the electron correlation effects 
using the above many-body methods that are restricted at different levels of approximations and 
to demonstrate furtherance in the preciseness of the results by carrying out large scale computations 
involved in some of these employed methods. We apply these methods to determine polarizabilities of B$^+$, C$^{2+}$, Al$^+$, Si$^{+2}$, Zn, Ga$^+$, 
Ge$^{+2}$, Cd, In$^+$ and Sn$^{+2}$, those belong to the transition and post-transition metallic groups
of the periodic table. We also explicitly investigate the contributions arising through the non-linear
mathematical expressions constituting the higher order excitation processes by setting-up 
intermediate maneuver to curtail the computational time at the expense of large memory 
requirements for the goal of promoting accuracies in the results compared to the available measurements.

The rest of the paper is organized as follows: In the next section we give briefly the theory of the atomic dipole 
polarizability. In section \ref{sec3} we describe many-body methods at different levels of approximations. Before
concluding the present work, we give our results in section \ref{sec4} and compare them with the 
other available calculations and measurements. Unless stated otherwise atomic unites are considered 
throughout this paper.

\section{Theory of Dipole Polarizability}\label{sec2}

The change in energy of the ground state in an atomic system due to the application of an external electric
field $\vec{\mathcal E}$ is given by
\begin{eqnarray}
 \Delta E=-\frac{1}{2}\alpha |\vec{\mathcal E}|^2,
 \label{eq1}
\end{eqnarray}
where $\alpha$ is known as the dipole polarizability of the state. In the mathematical
expression, we can write
\begin{equation}
\alpha=- 2 \frac{\langle \Psi_0^{(0)}|D|\Psi_0^{(1)} \rangle}{ \langle \Psi_0^{(0)}| \Psi_0^{(0)} \rangle },
\label{eq3}
\end{equation}
with $|\Psi_0^{(0)} \rangle$ and $|\Psi_0^{(1)} \rangle$ are the unperturbed and the first-order perturbed ground
state wave functions due to the interaction Hamiltonian $\vec D . \vec{\mathcal E}$ due to the dipole operator $D$. The arduous
part of calculating $\alpha$ using the above expression lies in the evaluation of $|\Psi_0^{(1)} \rangle$
which entails mixing of different parity states. On the other hand, it is sometimes facile to use a
sum-over-states approach given by
\begin{equation}
\alpha=-\frac{2}{\langle \Psi_0^{(0)}| \Psi_0^{(0)} \rangle } \sum_{I}\frac{|\langle \Psi_0^{(0)}|D|\Psi_I^{(0)} \rangle|^2}{E_0^{(0)}-E_I^{(0)}},
\label{eq2}
\end{equation}
where $I$ represents all possible intermediate states $|\Psi_I^{(0)} \rangle$ and $E_K^{(0)}$s are the 
energies of the respective $K$ states denoted by the indices in the subscripts. The above approach is 
convenient to use if the electric dipole (E1) matrix elements between the ground state and a sufficient number of 
intermediate states are known or can be calculated to the reasonable accuracies. However, it is extremely
difficult to determine these matrix elements accurately with confidence as it requires careful handling of a large number of 
configuration state functions (CSFs). Moreover, contributions coming from the core, doubly excited states,
continuum etc. cannot be accounted correctly through a sum-over-states approach and estimating these 
contributions approximately may be an extortionate practice at times when the systems are almost quasi-degenerate
in nature.

 The other famous approach for determining $\alpha$ is using the finite $\vec{\mathcal E}$ perturbation
method in which the second order differentiation of the total energy ($E_0$) of the ground state need to
be estimated in the presence of the electric field (finite field method); i.e.
\begin{equation}
 \alpha= - \left ( \frac{ \partial^2 E_0 (|\vec{\mathcal E}|)}{\partial |\vec{\mathcal E}| \partial |\vec{\mathcal E}|} \right )_{|\vec{\mathcal E}|=0},
\end{equation}
which requires numerical calculations for a smaller arbitrary value of $\vec{\mathcal E}$. This is a
typical procedure of calculating $\alpha$ using the molecular methods based on the Cartesian coordinate
systems where the atomic states do not possess definite parity. In contrast, it is a convoluted 
procedure of determining $\alpha$ of the atomic systems in the relativistic formalism if we wish to describe 
the method exclusively in the spherical coordinates. 

 Our methodology to determine $\alpha$ lies in the technique of calculating $|\Psi_0^{(1)} \rangle$ and to
supplant the ideology of obtaining it as the solution of the following inhomogeneous equation
\begin{eqnarray}
(H-E_0^{(0)}) |\Psi_0^{(1)} \rangle &=& -D|\Psi_0^{(0)} \rangle
\label{eq4}
\end{eqnarray}
through the matrix mechanism in the four-component relativistic theory described by the spherical 
polar coordinate system. Approximating the total wave function of the ground state to
$|\Psi_0 \rangle \simeq |\Psi_0^{(0)} \rangle + \lambda |\Psi_0^{(1)} \rangle$, we have 
\begin{equation}
 \alpha= \frac{\langle \Psi_0|D|\Psi_0 \rangle}{\langle \Psi_0|\Psi_0 \rangle},
\label{eq6}
\end{equation}
where $\lambda$ is an arbitrary parameter to identify the order of perturbation in $D$.

\section{Few Many-body methods for $ |\Psi_0^{(1)} \rangle$}\label{sec3}
We consider the DC atomic Hamiltonian in our calculations that is scaled with respect to the rest mass energies
of the electrons and is given by
\begin{eqnarray}
\nonumber H^{DC}=\sum_i \left [ c\mbox{\boldmath$\alpha$}_i\cdot \textbf{p}_i+(\beta_i -1)c^2+
V_{nuc}(r_i) + \sum_{j>i} \frac{1}{r_{ij}} \right ] \\ &&
\label{eq7}
\end{eqnarray}
where $c$ is the velocity of light, $\mbox{\boldmath$\alpha$}$ and $\beta$ are the Dirac matrices in their
fundamental representations, $r_{ij}$ represent the inter-electronic distances
and $V_{nuc}(r)$ is the nuclear potential calculated by considering the finite-size nuclear Fermi charge
distribution as given by
\begin{eqnarray}
 \rho_{nuc}(r)=\frac{\rho_0}{1+e^{(r-c)/a}}
 \label{eq8}
\end{eqnarray}
with the parameter $c$ and $a=4 t ln(3)$ are said to be the half-charge-radius and skin thickness of 
the nucleus, respectively.

We determine the approximated wave function ($|\Phi_0 \rangle$) for the ground state using the 
mean-field method by defining the Dirac-Fock (DF) Hamiltonian as
\begin{eqnarray}
H_{DF} &=& \sum_i  [c\mbox{\boldmath$\alpha$}_i\cdot \textbf{p}_i+(\beta_i -1)c^2+  
V_{nuc}(r_i) + U_{DF}(r_i)] \nonumber \\
      &=& \sum_i [h_0(r_i) + U_{DF}(r_i)],
\label{eq9}
\end{eqnarray}
with an average DF potential $U_{DF}(r)$ and disregarding contributions from the residual interaction
\begin{eqnarray}
 V_{es}=\sum_{j>i}^N \frac{1}{\vec{r}_{ij}} -\sum_i U_{DF}(\vec{r}_i).
 \label{eq10}
\end{eqnarray}
The DF potential and the single particle wave function $|\phi_i^0 \rangle$ of $|\Phi_0 \rangle$
are obtained by solving the following equations
\begin{eqnarray}
\nonumber \langle \phi_i^0|U_{DF}|\phi_j^0 \rangle=\sum_b^{occ}[\langle \phi_i^0 \phi_b^0|\frac{1}{r_{12}} |\phi_b^0\phi_j^0\rangle
-\langle \phi_i^0 \phi_b^0|\frac{1}{r_{12}} |\phi_j^0 \phi_b^0\rangle]\\
\label{eq13}
\end{eqnarray}
and
\begin{eqnarray}
 (h_0+U_{DF})|\phi_i^0\rangle &=& \epsilon_i^0|\phi_i^0\rangle
 \label{eq12}
\end{eqnarray}
simultaneously in a self-consistent procedure, where $b$ is summed over all the occupied orbitals ($occ$).

To get the wave function $|\Psi_0^{(0)} \rangle$, we follow the Bloch equation formalism \cite{lindgren} in which 
we express
\begin{eqnarray}
|\Psi_0^{(0)} \rangle &=&  \Omega^{(0)} |\Phi_0 \rangle \nonumber \\
                      &=& \sum_k^n \Omega^{(k,0)} |\Phi_0 \rangle,
\end{eqnarray}
where $\Omega^{(0)}$ is known as wave operator containing $n$ (say) orders of Coulomb interactions and $k$ represents order of $V_{es}$ in the wave operator
in a perturbative expansion of $\Omega^{(0)}$. In the presence of another external interaction, like the operator $D$,
the exact state can be written as
\begin{eqnarray}
 |\Psi_0 \rangle &=& \Omega |\Phi_0 \rangle \nonumber \\
                 &=& \sum_{\beta}^n \sum_{\delta}^m \Omega^{(\beta,\delta)} |\Phi_0 \rangle,
 \label{eq14}
\end{eqnarray}
where the perturbation expansion is again described by $n$ (say) orders of $V_{es}$ and $m$ (say) orders
of $D$. For our requirement of obtaining the first order wave function due to $D$, we have
\begin{eqnarray}
 |\Psi_0^{(1)} \rangle  &=& \sum_{\beta}^n \Omega^{(\beta,1)} |\Phi_0 \rangle .
\end{eqnarray}

To obtain the solutions for the wave operators, we use the following generalized Bloch equations
\begin{eqnarray}
 [\Omega^{(\beta,0)},H_{DF} ] P &=& Q V_{es} \Omega^{(\beta-1,0)}P  \nonumber \\ && -
 \sum_{m=1 }^{\beta-1} \Omega^{(\beta-m,0)} P V_{es} \Omega^{(m-1,l)}P \ \  \  \  \
\end{eqnarray}
and
\begin{eqnarray}
 [\Omega^{(\beta,1)},H_{DF} ]P &=& QV_{es} \Omega^{(\beta-1,1)}P + Q D \Omega^{(\beta,0)}P 
\nonumber \\ && - \sum_{m=1 }^{\beta-1} \big ( \Omega^{(\beta-m,1)}
 P V_{es} \Omega^{(m-1,0)}P \nonumber \\ && - \Omega^{(\beta-m,1)}PD \Omega^{(m,0)}P \big ),
\label{eq15}
\end{eqnarray}
with the definitions of $P= |\Phi_0 \rangle \langle \Phi_0 |$ and $Q=1-P$. It implies that 
$\Omega^{(0,0)}=1$, $\Omega^{(1,0)}= \sum_{p,a} \frac{ \langle \Phi_a^p | V_{es} | \Phi_0 \rangle} 
{ E_a^p - E_0} = 0$ and $\Omega^{(0,1)}= \sum_{p,a} \frac{ \langle \Phi_a^p | D | \Phi_0 \rangle} 
{ E_a^p - E_0} = \sum_{p,a} \frac{ \langle p | D | a \rangle} { \epsilon_p^0 - \epsilon_a^0}$ with 
$a$ and $p$ representing the occupied and unoccupied orbitals, respectively.

 Below we discuss various many-body methods in the DF, MBPT(3), RPA and CC approaches employed in the
present work based on the above discussed Bloch's equation formalism to calculate $\alpha$s in the considered 
atomic systems for their case studies. Among these methods, we have already described the DF, MBPT(3) and CC 
methods elaborately and given the final $\alpha$ evaluating diagrams for the closed-shell atomic systems in our 
earlier work \cite{yashpal-polz}. For the sake of completeness we would like to outline these methods here,
but describe the RPA method extensively.

\subsection{The DF method}

 The lowest order polarizabilities results in the DF method are evaluated by using the expression
\begin{eqnarray}
 \alpha  &=& 2 \langle \Phi_0| {\Omega^{(0,0)}}^{\dagger} D \Omega^{(0,1)} |\Phi_0 \rangle \nonumber \\
         &=& 2 \langle \Phi_0| D \Omega^{(0,1)} |\Phi_0 \rangle .
\end{eqnarray}

\subsection{The MBPT(3) method}

In this  approximation, we have considered two orders of Coulomb ($\beta=2$) for which the expression for $\alpha$
is given by 
\begin{eqnarray}
\alpha &=& 2 \frac{\sum_{\beta=0}^{2} \langle \Phi_0| {\Omega^{(2-\beta,0)}}^{\dagger} D \Omega^{(\beta,1)} |\Phi_0 \rangle}
{ \sum_{\beta=0}^{2} \langle \Phi_0| {\Omega^{(2-\beta,0)}}^{\dagger} \Omega^{(\beta,0)} |\Phi_0 \rangle} \nonumber \\
       &=& \frac{2}{\cal N} \langle \Phi_0 | [\Omega^{(0,0)}+\Omega^{(1,0)}+\Omega^{(2,0)}]^{\dagger} D \nonumber \\ && \times[\Omega^{(0,1)}+\Omega^{(1,1)}+\Omega^{(2,1)}]|\Phi_0 \rangle \nonumber \\
&=& \frac{2}{\cal N} \langle \Phi_0| D\Omega^{(0,1)} + D\Omega^{(1,1)}+D\Omega^{(2,1)} + {\Omega^{(1,0)}}^{\dagger} D\Omega^{(0,1)}  \nonumber \\ && +
{\Omega^{(1,0)}}^{\dagger} D\Omega^{(1,1)} +{\Omega^{(2,0)}}^{\dagger} D\Omega^{(0,1)}|\Phi_0 \rangle  ,
\label{eq21}
\end{eqnarray} 
with the normalization constant ${\cal N}=\langle \Phi_0| 1 + \Omega^{(1,0)}+\Omega^{(2,0)} + {\Omega^{(1,0)}}^{\dagger} + 
{\Omega^{(2,0)}}^{\dagger} +{\Omega^{(1,0)}}^{\dagger} \Omega^{(0,1)} |\Phi_0 \rangle$. This clearly
means that its lowest order corresponds to the MBPT(1) method which is nothing but the DF contribution. Terms
containing up to one order of Coulomb and one $D$ operator is referred to the MBPT(2) method.

\subsection{The RPA method}

\begin{figure}[t]
\includegraphics[width=8cm]{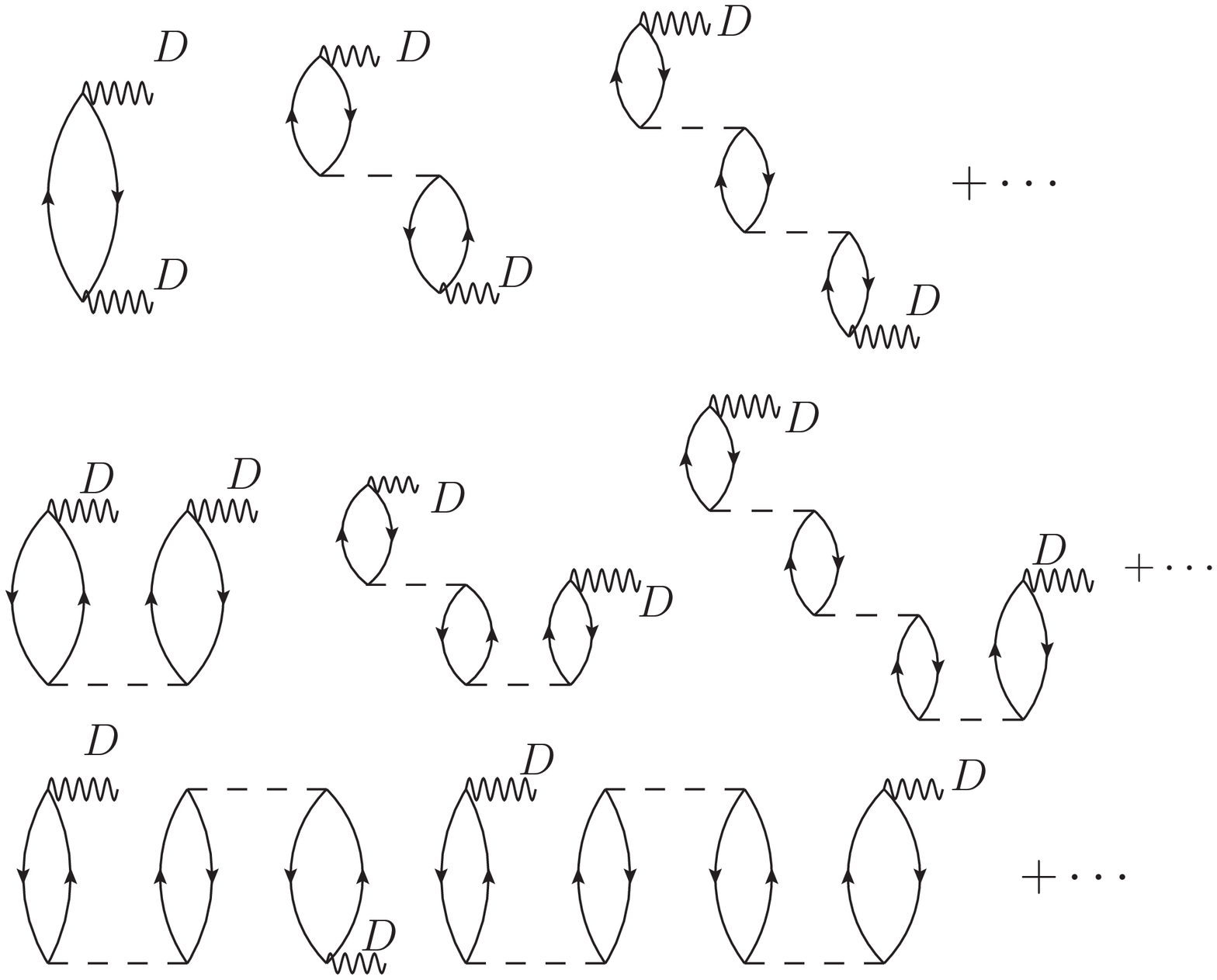}
\caption{Dominant direct core-polarization diagrams contributing to the $\alpha$ calculations in the RPA method.}
 \label{fig1}
\end{figure}

To arrive at the final working equation for the RPA method, we start by perturbing the DF orbitals 
and single particle energies due to the perturbation $D$. i.e. 
\begin{eqnarray}
  |\phi_i^0\rangle && \rightarrow |\phi_i^0\rangle+\lambda|\phi_i^1\rangle \\
 \text{and} \ \ \ \ \epsilon_i^0 &&\rightarrow \epsilon_i^0 + \lambda \epsilon_i^1 ,
 \label{eq22}
\end{eqnarray}
where $|\phi_i^1\rangle$ and $\epsilon_i^1$ are the first order corrections to the particle wave function 
and energy, respectively. Owing to the fact that $D$ is an odd parity operator, $\epsilon_i^1=0$.
Now in the presence of the perturbative source, the modified DF equation for the single particle 
wave function yields the form
\begin{eqnarray}
 \nonumber (h_0+\lambda D)(|\phi_i^0 \rangle+\lambda|\phi_i^1\rangle)+  \sum_b^{occ} (\langle \phi_b^0+ \lambda \phi_b^1 |
 \frac{1}{r_{12}} |\phi_b^0\\ \nonumber  + \lambda \phi_b^1 \rangle |\phi_i^0 +\lambda \phi_i^1\rangle -
 \langle \phi_b^0+ \lambda  \phi_b^1 |\frac{1}{r_{12}} |\phi_i^0+ \lambda \phi_i^1 \rangle |\phi_b^0 \\ +\lambda \phi_b^1\rangle) 
  = \epsilon_i^0(|\phi_i^0 \rangle+\lambda|\phi_i^1\rangle).
 \label{eq23}
\end{eqnarray}
Collecting the terms only those are linear in $\lambda$, we get
\begin{eqnarray}
 (h_0+U_{DF}-\epsilon_i^0)|\phi_i^1\rangle= (-D -U_{DF}^1)|\phi_i^0\rangle,  
 \label{eq24}
\end{eqnarray}
where we use the notation $U_{DF}^1$ for
\begin{eqnarray}
 \nonumber U_{DF}^1 |\phi_i^0\rangle =\sum_b^{occ} (\langle \phi_b^0|\frac{1}{r_{12}} |\phi_b^1\rangle |\phi_i^0 \rangle 
 -\langle \phi_b^0|\frac{1}{r_{12}} |\phi_i^0\rangle |\phi_b^1 \rangle \\ +\langle \phi_b^1|\frac{1}{r_{12}} |\phi_b^0\rangle |\phi_i^0 \rangle
 -\langle \phi_b^1|\frac{1}{r_{12}} |\phi_i^0\rangle |\phi_b^0 \rangle) . \  \  \  \
 \label{eq25}
\end{eqnarray}

  Following basic principles we can write the single particle perturbed wave function in terms of
  unperturbed wave functions as
\begin{eqnarray}
 |\phi_i^1\rangle=\sum_{j \ne i} C_i^j |\phi_j^0\rangle,
 \label{eq26}
\end{eqnarray}
where $C_i^j$s are the expansion coefficients. In the RPA approach, we write
\begin{eqnarray}
 \sum_{j \ne i} C_i^i (h_0 + U_{DF} - \epsilon_j^0 ) |\phi_j^0\rangle= (- D - U_{DF}^1) |\phi_i^0\rangle,  
\end{eqnarray}
and solve this equation self-consistently to obtain the $C_i^j$ coefficients with infinity orders
of contributions from the Coulomb interaction considering their initial solutions as the above perturbed DF method. 

In the Bloch's wave operator representation, we can express
\begin{eqnarray}
\Omega_{\text{RPA}}^{(1)} &=&  \sum_k^{\infty} \sum_{p,a} \Omega_{a \rightarrow p}^{(k, 1)} \nonumber \\
    &=& \sum_{\beta=1}^{\infty} \sum_{pq,ab} { \{} \frac{[\langle pb | \frac{1}{r_{12}} | aq \rangle 
- \langle pb | \frac{1}{r_{12}} | qa \rangle] \Omega_{b \rightarrow q}^{(\beta-1,1)} } {\epsilon_p - \epsilon_a}  \nonumber \\ 
&& + \frac{ \Omega_{b \rightarrow q}^{{(\beta-1,1)}^{\dagger}}[\langle pq | \frac{1}{r_{12}} | ab \rangle - \langle pq | \frac{1}{r_{12}} | ba \rangle] 
}{\epsilon_p-\epsilon_a} { \}},
\label{eq27}
\end{eqnarray} 
where $a \rightarrow p$ means replacement of an occupied orbital $a$ from $|\Phi_0 \rangle$ by a
virtual orbital $p$ which alternatively refers to a singly excited state with respect to $|\Phi_0 \rangle$.
It can be shown from the above formulation that the RPA method picks-up a certain class of singly excited
configurations congregating the core-polarization correlation effects to all orders. 

Using the above RPA wave operator, we evaluate $\alpha$ by
\begin{eqnarray}
 \alpha  &=& 2 \langle \Phi_0| {\Omega^{(0,0)}}^{\dagger} D \Omega_{\text{RPA}}^{(1)} |\Phi_0 \rangle \nonumber \\
         &=& 2 \langle \Phi_0| D \Omega_{\text{RPA}}^{(1)} |\Phi_0 \rangle .
\end{eqnarray}
Impediment of this method is that it encapsulates contributions to $|\Psi_0^{(1)} \rangle$ from the
correlation effects due to the Coulomb interaction to all orders, but only from the core-polarization effects 
through the singly excited configurations. However, it approximates the bra state $|\Psi_0^{(0)} \rangle$
of Eq. (\ref{eq3}) to the mean-field wave function $|\Phi_0 \rangle$. Diagrammatic representation of the 
core-polarization correlations embraced through RPA are given (without the exchange interactions) in 
Fig. \ref{fig1}. 

\subsection{The CC method}
\begin{figure}[t]
 \includegraphics[width=7.5cm]{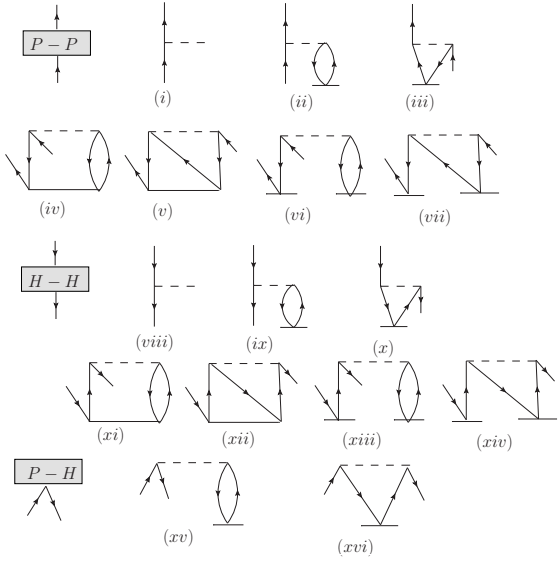}
 \caption{Effective one-body intermediate diagrams constructed from $\overline{H_N^{DC}}$ in order to evaluate the $T^{(1)}$ amplitudes.
 Here broken lines represent the Coulomb interaction and the solid line without arrows symbolize the $T^{(0)}$ operators.}
 \label{oneb}
\end{figure}

\begin{figure}[t]
 \includegraphics[width=7.4cm]{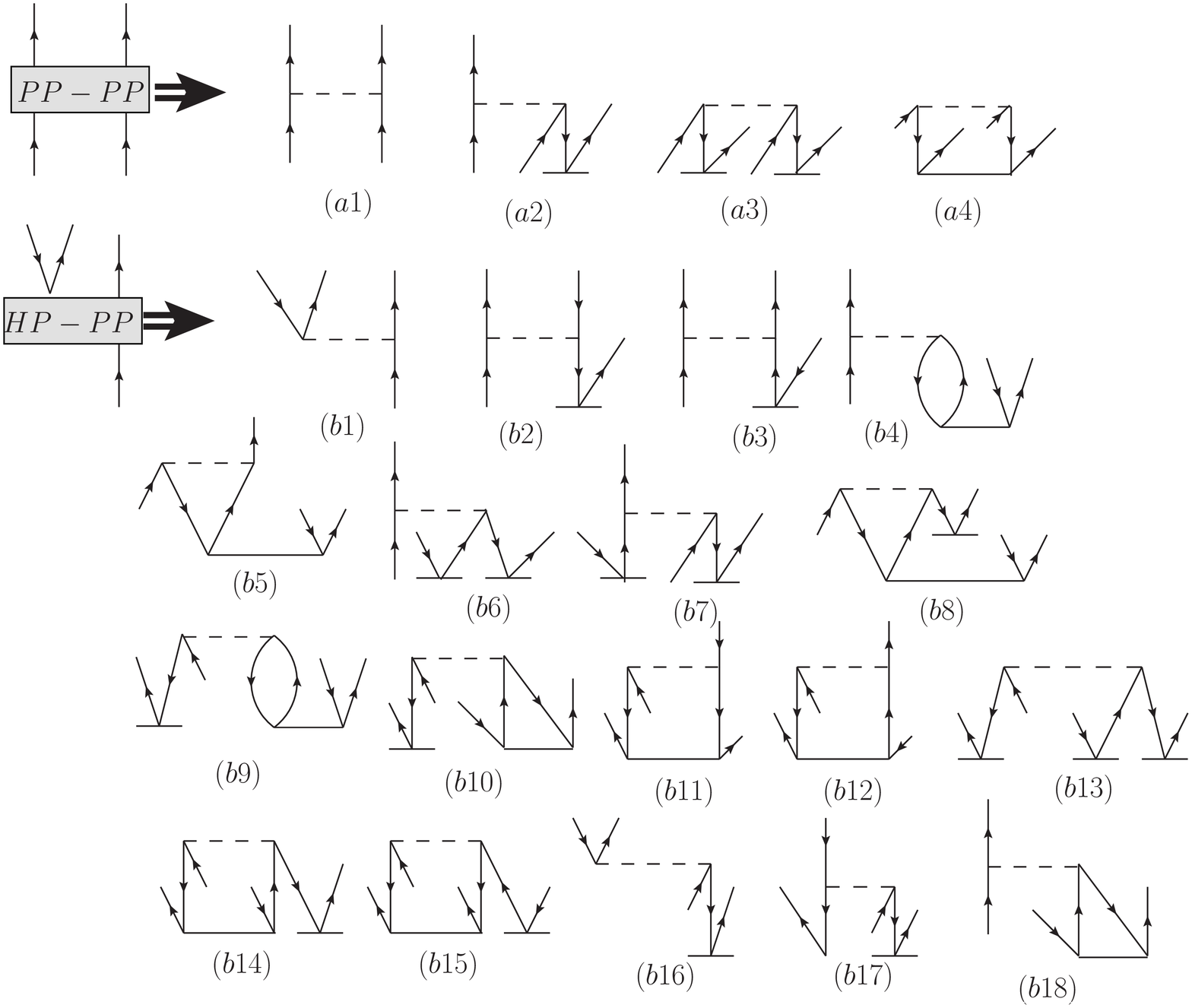}
 \includegraphics[width=7.4cm]{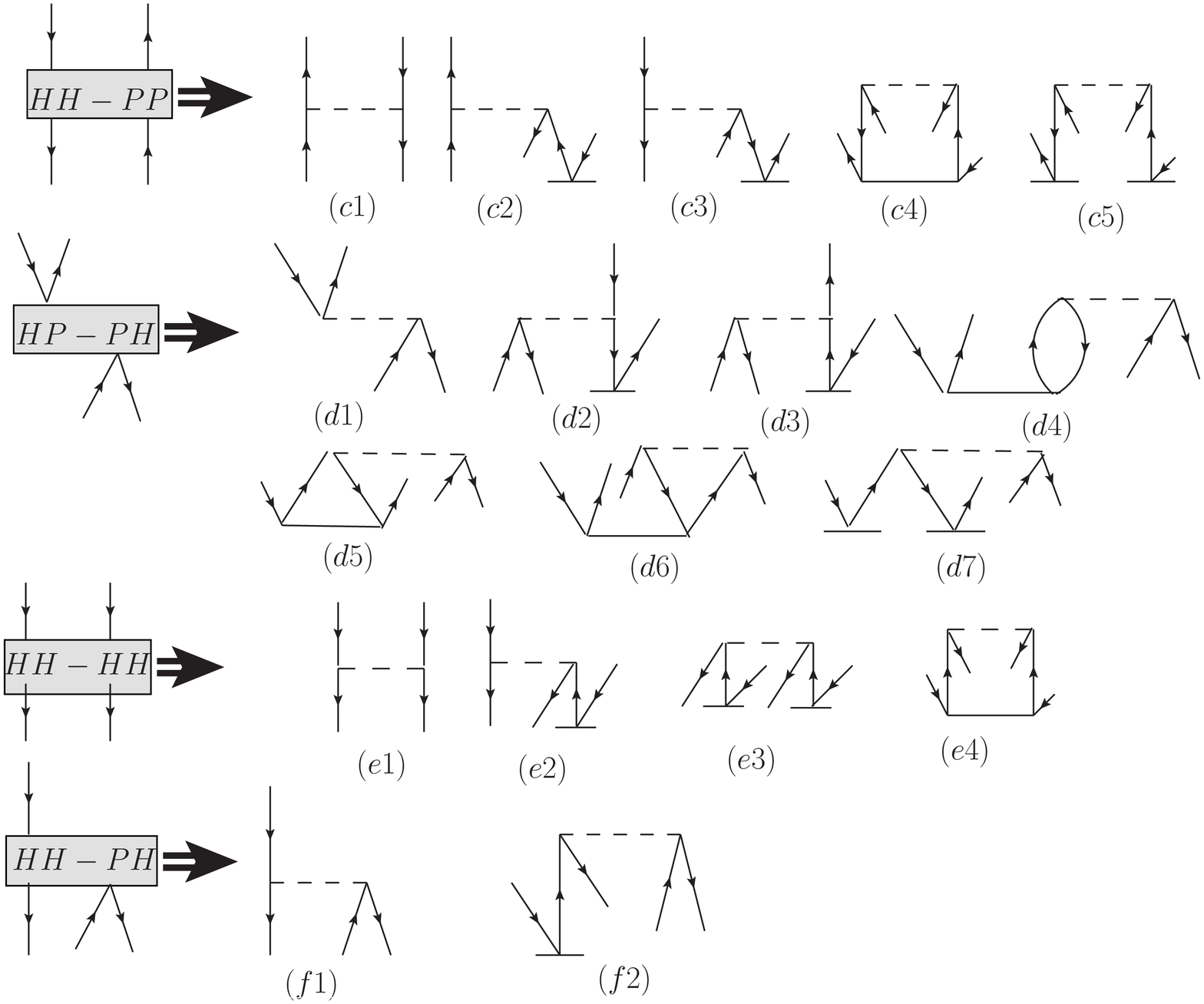}
 \includegraphics[width=7.3cm]{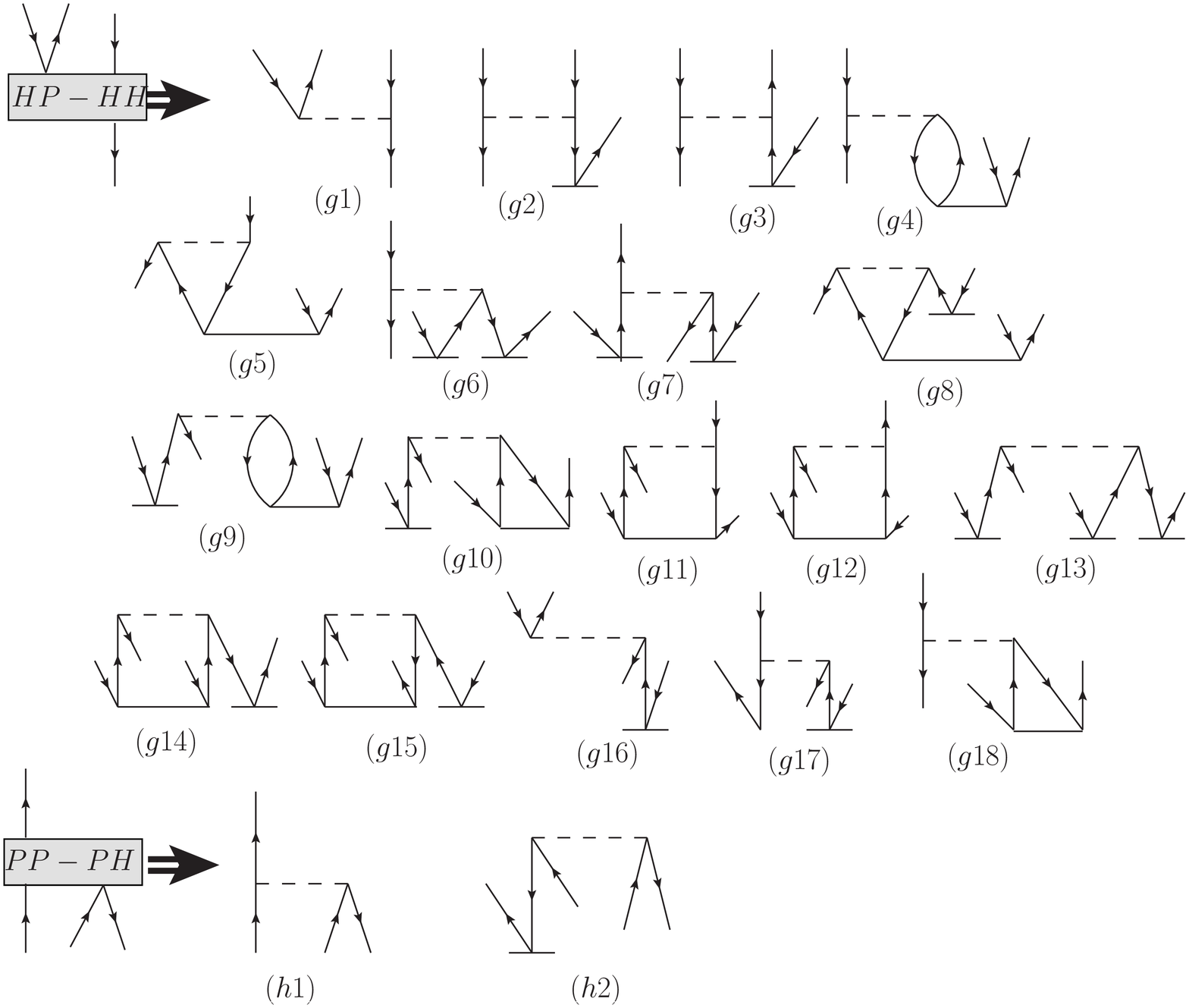}
 \caption{Effective two-body intermediate diagrams constructed from $\overline{H_N^{DC}}$ in order to evaluate the $T^{(1)}$ amplitudes.
 Here broken lines represent the Coulomb interaction and the solid line without arrows symbolize the $T^{(0)}$ operators.}
 \label{twob}
\end{figure}

\begin{figure}[t]
 \includegraphics[width=7.5cm]{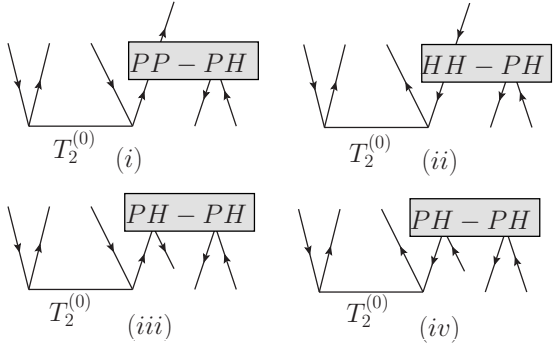}
 \caption{Effective three-body intermediate diagrams constructed from $\overline{H_N^{DC}} T_2^{(0)}$ to solve for 
 the $T^{(1)}$ amplitudes.}
 \label{tbody}
\end{figure}
In the CC method, we express the unperturbed atomic wave function as
\begin{eqnarray}
 | \Psi_0^{(0)} \rangle &=& \Omega_{\text{RCC}}^{(0)}|\Phi_0 \rangle  \nonumber \\ 
                        &=& \sum_k^{\infty} \Omega_{\text{RCC}}^{(k,0)}|\Phi_0 \rangle  \nonumber \\ 
                        &=& e^{T^{(0)}} |\Phi_0 \rangle 
\label{eq32}
\end{eqnarray}
and the first order perturbed wave function as \cite{yashpal-polz}
\begin{eqnarray}
 | \Psi_0^{(1)} \rangle &=& \Omega_{\text{RCC}}^{(1)} |\Phi_0 \rangle \nonumber \\ 
                        &=& \sum_k^{\infty} \Omega_{\text{RCC}}^{(k,1)}|\Phi_0 \rangle  \nonumber \\ 
                        &=& e^{T^{(0)}} T^{(1)} |\Phi_0 \rangle, 
\label{eq33}
\end{eqnarray} 
where $T^{(0)}$ and $T^{(1)}$ are the excitation operators from the reference state $|\Phi_0 \rangle$
that take care of contributions from the Coulomb interactions and Coulomb interactions along with from the 
perturbed $D$ operator, respectively. 

The amplitudes of the excitation $T^{(0)}$ and $T^{(1)}$ operators are determined using the equations
\begin{eqnarray}
 \langle \Phi_0^{\tau}|\overline{H_N^{DC}}|\Phi_0\rangle&=&0
 \label{eq36}
 \end{eqnarray}
and
\begin{eqnarray}
\langle \Phi_0^{\tau}|\overline{H_N^{DC}}T^{(1)}|\Phi_0\rangle&=&-\langle \Phi_0^{\tau}|\overline{D}|\Phi_0\rangle ,
  \label{eq37}
\end{eqnarray}
where $H_N^{DC}$ is the normal ordered DC Hamiltonian, $\overline{O}=(Oe^{T^{(0)}})_{con}$ with $con$
means only the connected terms and $| \Phi_0^{\tau} \rangle$ corresponds to the 
excited configurations with $\tau$ referring to level of excitations from $| \Phi_0 \rangle$. In our 
calculations, we only consider the singly and doubly excited configurations ($\tau=1,2$) by defining
\begin{eqnarray}
T^{(0)} &=& T_1^{(0)} + T_2^{(0)} \ \ \ \ \text{and} \ \ \ \ 
T^{(1)} = T_1^{(1)} + T_2^{(1)} ,
\label{eq35}
\end{eqnarray}
which is known as the CCSD method in the literature. When we consider the approximation 
$\overline{O}=O + O T$, we refer it as the LCCSD method.

To carry out calculations in an optimum computational requirements, we construct the intermediate diagrams 
for the effective operators by dividing the non-linear CC terms. The intermediate diagrams for the computation 
of the $T^{(0)}$ amplitudes are described at length in our previous work \cite{yashpal-polz}. Here, we discuss only about the 
intermediate diagrams used for the evaluation of the $T^{(1)}$ amplitudes. For this purpose, we express
$\overline{H_N^{DC}}$ into the effective one-body, two-body and three-body diagrams. It is worth while
to note that there is a technical difference between the construction of the intermediate diagrams from
$\overline{H_N^{DC}}$ for the $T^{(0)}$ and $T^{(1)}$ amplitude solving equations. In Eq. (\ref{eq37}),
$\overline{H_N^{DC}}$ contains all the non-linear terms while for solving Eq. (\ref{eq36}) it is
required to express as $\overline{H_N^{DC}}=\overline{H_N^{DC}}' \otimes T_{\tau}$. Thus the intermediate
diagrams in this case are comprised terms from $\overline{H_N^{DC}}'$ which requires special scrutiny of
the diagrams to avoid repetition in the singles and doubles amplitude calculations. The effective intermediate 
diagrams used for the $T^{(1)}$ amplitude determining equations are shown in Figs. \ref{oneb}, \ref{twob} 
and \ref{tbody}. These effective diagrams are finally connected with the respective $T^{(1)}$ operators to 
obtain the amplitudes of the singles and doubles excitations and the corresponding diagrams are presented 
in Figs. \ref{sing-con} and \ref{doub-con}. Contributions from the terms of $\overline{D}$ are evaluated 
directly for the $T^{(1)}$ amplitude calculations and the corresponding diagrams are shown in Figs. 
\ref{dbar1} and \ref{dbar2}. 

In order to estimate the dominant contributions from the triple excited configurations, we define
an excitation operator perturbatively in the CC framework as following
\begin{eqnarray}
 T_3^{(0),pert}= \frac{1}{3!}\sum_{abc}^{pqr} 
 \frac{ ( \overline{H_N^{DC}} T_2^{(0)})_{abc}^{pqr} }{\epsilon_a
 + \epsilon_b+\epsilon_c-\epsilon_p -\epsilon_q -\epsilon_r} 
 \label{eq30}
\end{eqnarray}
\begin{figure}[t]
 \includegraphics[width=7cm]{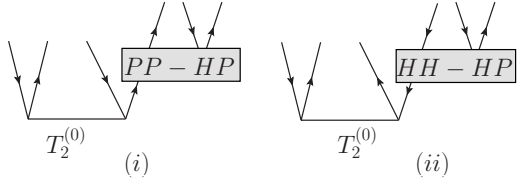}
 \caption{Diagrams representing $T_3^{(0),pert}$ operator.}
 \label{ptrip}
\end{figure}
which diagrammatically shown in Fig. \ref{ptrip} and contract it with the $D$ operator to calculate 
the amplitudes of the $T_2^{(1)}$ perturbed CC operator in a self-consistent procedure
considering it in Eq. (\ref{eq37}) as part of $\overline{D}$. We refer this
approach as the CCSD$_p$T method in this work.

\begin{figure}[t]
 \includegraphics[width=7.0cm]{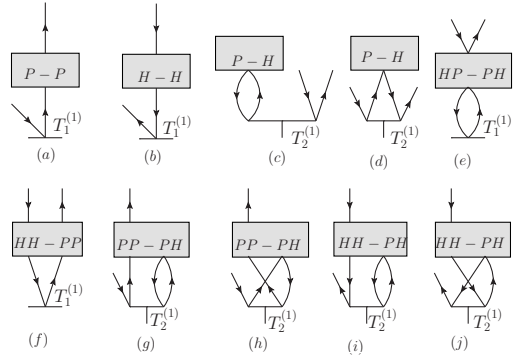}
 \caption{Final contributing diagrams for the $T_1^{(1)}$ amplitude calculations which are constructed by the contraction of effective one and two
 body intermediate diagrams with $T^{(1)}1$ operators.}
 \label{sing-con}
\end{figure}

\begin{figure}[t]
 \includegraphics[width=7.0cm]{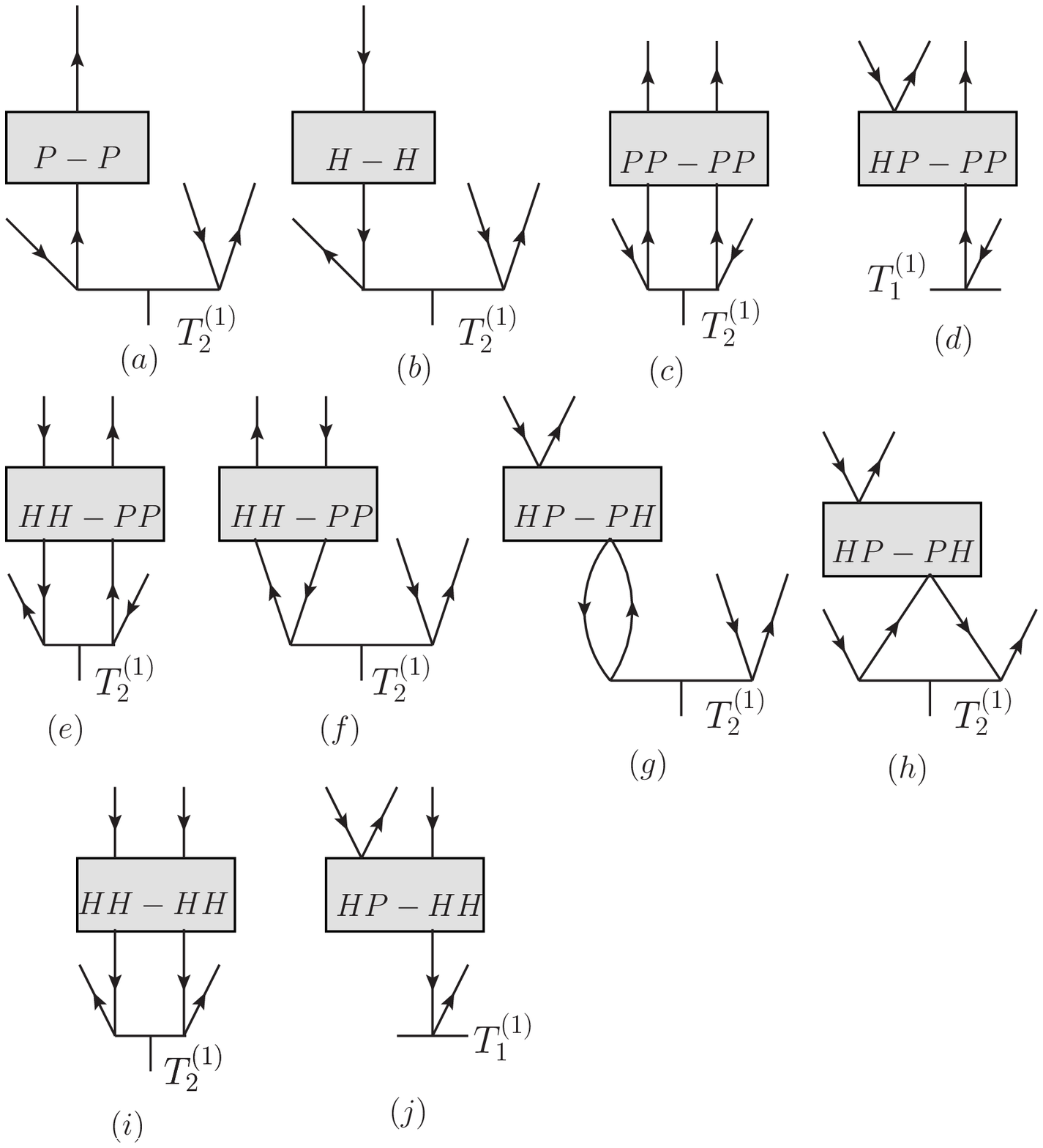}
 \caption{Final contributing diagrams for the $T_2^{(1)}$ amplitude calculations which are constructed by the contraction of effective one and two
 body intermediate diagrams with $T^{(1)}1$ operators.}
 \label{doub-con}
\end{figure}

\begin{figure}[t]
 \includegraphics[width=7cm]{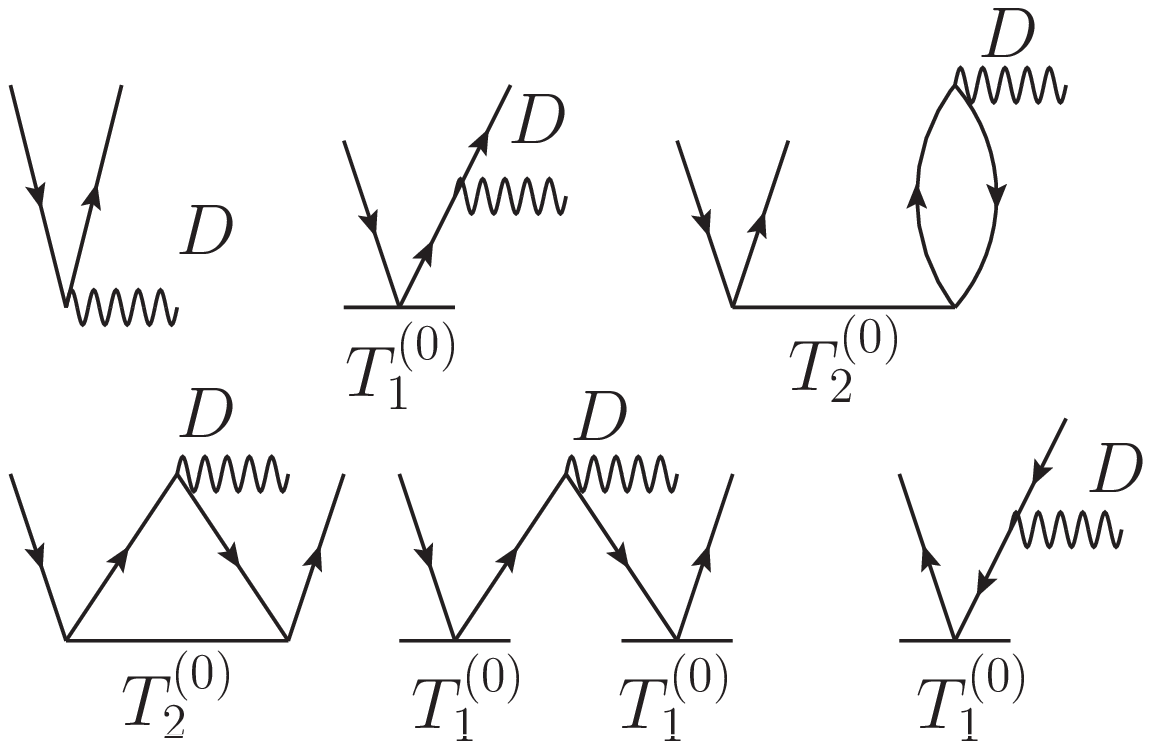}
 \caption{Single excitation diagrams from $\overline{D}$ that contribute to the calculations of the $T^{(1)}$ amplitudes.}
 \label{dbar1}
\end{figure}

\begin{figure}[t]
 \includegraphics[width=7cm]{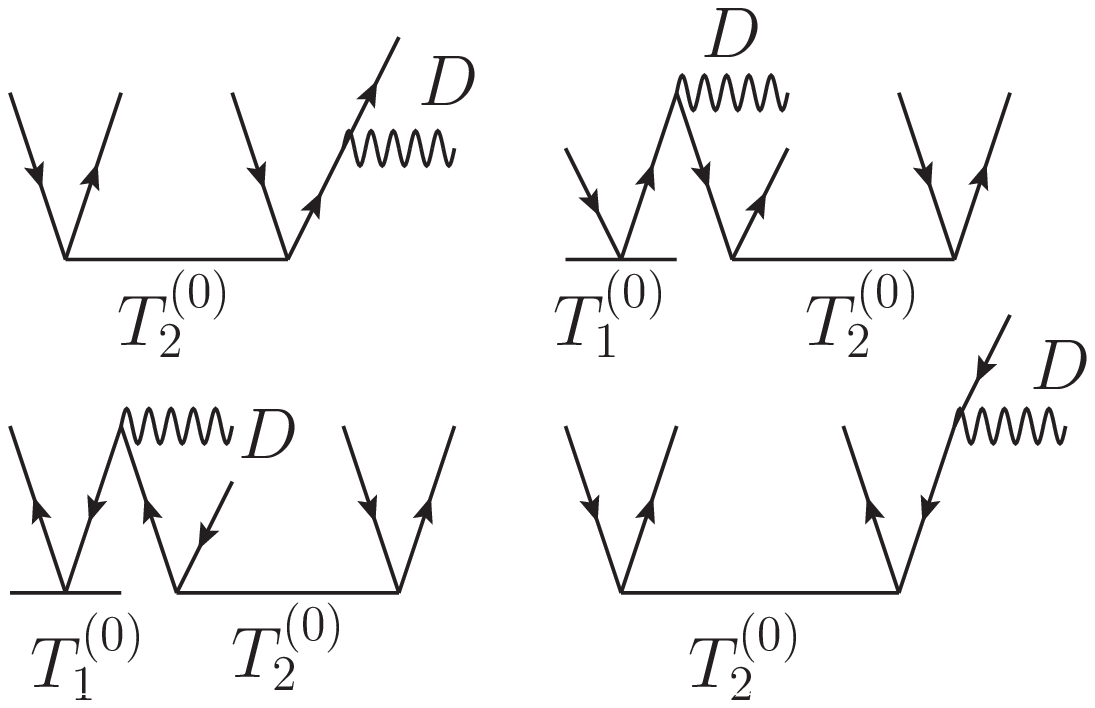}
 \caption{Double excitation diagrams from $\overline{D}$ that contribute to the calculations of the $T^{(1)}$ amplitudes.}
 \label{dbar2}
\end{figure}

\begin{figure}[t]
 \includegraphics[width=8cm]{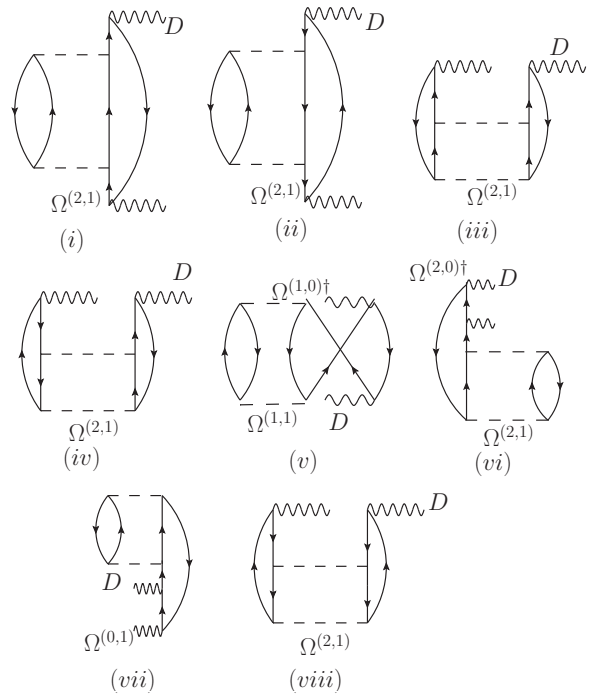}
 \caption{Few significantly contributing non-RPA type MBPT(3) diagrams.}
 \label{mbpt}
\end{figure}

Using the above formulation, the expression for the polarizability is given by \cite{yashpal-polz}
\begin{eqnarray}
 \alpha &=&2 \frac{\langle\Phi_0 | \Omega_{\text{RCC}}^{(0) \dagger} D \Omega_{\text{RCC}}^{(1)} | \Phi_0 \rangle }
                  {\langle\Phi_0 | \Omega_{\text{RCC}}^{(0) \dagger} \Omega_{\text{RCC}}^{(0)} | \Phi_0 \rangle } \nonumber \\
        &=&2 \frac{\langle\Phi_0 | e^{T^{(0) \dagger}} D e^{T^{(0)}} T^{(1)} | \Phi_0 \rangle }
                  {\langle\Phi_0 | e^{T^{(0) \dagger}} e^{T^{(0)}} | \Phi_0 \rangle } \nonumber \\
        &=&2 \langle\Phi_0 |(\overbrace{D^{(0)}} T^{(1)})_{con}|\Phi_0 \rangle,
\label{eq38}
\end{eqnarray}
where $\overbrace{D^{(0)}} = e^{T^{\dagger{(0)}}}De^{T^{(0)}}$ is a non-truncating series. In the LCCSD
method, we only consider the terms $\overbrace{D^{(0)}} = O + OT^{(0)} + T^{\dagger{(0)}} O + T^{\dagger{(0)}} O T^{(0)}$.
Computational steps to account all the significant contributions from $\overbrace{D_N^{(0)}}$ have
been described in detail in our previous work \cite{yashpal-polz}.

\section{Results and Discussion}\label{sec4}

\begin{table}[t]
\caption{Results for the dipole polarizabilities from our CCSD$_p$T method along with the available measurements and others calculations.
Uncertainties in the results are given in the parentheses and the references are cited in the square brackets.}
\begin{ruledtabular}
\begin{tabular}{lccc}
\textrm{System}& Present  &\multicolumn{2}{c}{\textrm{Others}} \\
               &  CCSD$_p$T  &  Theory & Experiment \\    
\hline                                                                       \\
B$^+$         & 10.395(20) & 9.448\cite{epstein}, 9.64(3)\cite{cheng1}  &\\
              &            & 9.624\cite{safronova}&\\
C$^{2+}$      & 4.244(10)  & 3.347\cite{epstein} &\\
Al$^+$        & 24.26(4)   & 24.2\cite{archibong}, 24.14(12)\cite{mitroy}& $^a$24.20(75)\cite{reshetnikov}\\
              &            & 24.12\cite{hamonou}, 24.048 \cite{safronova} &\\
              &            & 24.14(8) \cite{kallay} &\\
              &            & 24.065(410)\cite{yu} &\\
Si$^{2+}$     & 11.88(25)  & 11.688\cite{mitroy1} & 11.666(4)\cite{komara}\\
              &            & 11.75\cite{hamonou}& $^b$11.669(9)\cite{mitroy1}\\
Zn            & 38.666(35) & 38.12\cite{ye}, 39.2(8)\cite{goebel}& 38.8(8)\cite{goebel} \\
              &            & 38.4\cite{roos}, 37.9\cite{kello},  &  \\
              &            & 38.01\cite{seth} &  \\
Ga$^{+}$      & 18.441(20) & 17.95(34)\cite{cheng} & \\
Ge$^{2+}$     & 10.883(10) &  \\
Cd            & 45.856(42) &46.25\cite{seth}, 44.63\cite{ye}& 49.65(1.49)\cite{goebel1} \\
              &            &46.8\cite{kello}, 46.9\cite{roos}&\\
In$^{+}$      & 24.11(15)  & 24.01 \cite{safronova}& \\
Sn$^{2+}$     & 15.526(22)     &  \\
\end{tabular}
\end{ruledtabular}  
\label{tab1} 
\begin{tabular}{l}
$^a$ Estimated from the measured oscillator strengths.\\
$^b$ Obtained by reanalyzing data of Ref. \cite{komara}. \\
\end{tabular}
\end{table}  

Our final results using the CCSD$_p$T method along with the available experimental values for Al$^+$,
Si$^{2+}$, Zn and Cd and from the other calculations are given in Table \ref{tab1}. To ascertain lucidity
in the accuracies of the results from our calculations, we also provide the estimated uncertainties 
associated with our results by estimating contributions from various neglected sources and give them
in the parentheses alongside the CCSD$_p$T results in above table. The value that is referred to
as the experimental results for Al$^+$ is not directly obtained from the measurement, rather it is
estimated by summing over the experimental values of the oscillator strengths and has relatively large 
uncertainty compared to some of the reported calculations \cite{reshetnikov}. There are two high-precision 
results reported as the experimental values for the Si$^{2+}$ ion \cite{komara,mitroy1}, however the
value reported in \cite{komara} is obtained from the direct analysis of the energy intervals measurement
using the resonant Stark ionization spectroscopy (RESIS) technique while the other value \cite{mitroy1} is
reported by reanalyzing the data of Ref. \cite{komara}; which is about 0.03\% larger than the former value.
The only available experimental result of the ground state $\alpha$ of Zn is measured using an
interferometric technique by Goebel {\it et al.} \cite{goebel}. Similarly there is also one
measurement of $\alpha$ available for Cd using a technique of dispersive Fourier-transform 
spectroscopy, but the reported uncertainty in this experimental value is comparatively large \cite{goebel1}. 
Nevertheless when we compare our CCSD$_p$T results with all these experimental values, they match very well 
within their reported error bars except for Cd. In fact, our calculations are more precise in all 
the systems apart for the Si$^{2+}$ ion. There are no experimental results available for the other 
considered ions to compare them against our calculations.

There are also a number of calculations of $\alpha$ available by many groups using varieties
of many-body approaches among which some of them are based on either the lower order methods or
considering the non-relativistic mechanics. An old calculation of $\alpha$ in B$^+$ was reported
by Epstein {\it et al} \cite{epstein} based on the coupled perturbed Hartree-Fock (CHF) method while
Cheng {\it et al} had employed a configuration interaction method considering a semi-empirical
core-polarization potential (CICP method) to evaluate it more precisely \cite{cheng1}. Later 
Safronova {\it et al} used a combined CI and LCCSD methods (CI$+$all order method) to determine 
$\alpha$ of B$^+$ ion \cite{safronova}. However, the CCSD$_p$T result seems to
be larger than all other calculations. Our analysis suggests that the differences in these results are
mainly due to inclusion of the pair-correlation effects to all orders in our CC method. In C$^{2+}$ ion,
we find only one theoretical result reported by Epstein {\it et al} using the same CHF method.
Our result for C$^{+2}$ is also slightly larger than the value reported by the above calculation.
Till date Al$^+$ is the most precise ion clock in the world \cite{rosenband2} for which a couple of 
high-precision calculations have been reported on the determination of $\alpha$ of this ion by
attempting to push down the uncertainty in the black-body radiation (BBR) shift of the respective ion-clock
transition \cite{kallay,safronova,yu}. Among them calculations carried out by Mihaly {\it et al} is based
on the relativistic CC method considering up to quadrupole excitations and finite field approach
\cite{kallay}. However, calculations carried out in this work is based on the Cartesian coordinate
system and minimizing the energies in the numerical differentiation approach in contrast to the present CCSD$_p$T
method, where the matrix elements of $D$ are evaluated in the spherical coordinate system. Calculations
reported by Yu {\it et al} is using the same approach of Ref. \cite{kallay}, but by considering a different set 
of single particle orbitals \cite{yu}. Safronova {\it et al} have employed the CI$+$all order approach to
calculate $\alpha$ of Al$^+$. There are also other theoretical results have been reported based on
varieties of many-body methods such as CCSD, CICP, CI etc. both in the non-relativistic and relativistic
mechanics \cite{archibong,mitroy,hamonou}. We find an excellent agreement among
all the theoretical results. Some of these methods have also been employed to calculate $\alpha$ of
Si$^{2+}$ \cite{mitroy,hamonou} which are in perfect agreement with the experimental results. However,
our CCSD$_p$T value seems to be little larger then the experimental result but falls within the estimated uncertainty.
We found only one more calculation of $\alpha$ in Ga$^+$ using the CICP method
\cite{cheng} to compare with our result. Although values from both the calculations are very close but they 
do not agree within their reported uncertainties. Calculations in Cd are reported by many groups including 
the latest one using the Douglas-Kroll-Hess (DKH) Hamiltonian by Roos {\it et al} \cite{roos}. Calculations
carried out by Ye {\it et al} \cite{ye} is based on the relativistic formalism in the CICP method. All the theoretical
results are consistent and show good agreement with each other suggesting that the experimental result could have
been overestimated. Therefore it is important to have another measurement of the polarizability of Cd to resolve this ambiguity. 
Again there has also been an effort made for the precise determination of $\alpha$ in In$^+$
to estimate the BBR shift accurately for its proposed atomic clock transition \cite{safronova}. Our result
agrees nicely with this calculation. As discussed earlier, calculations carried out in \cite{safronova}
are based on the CI$+$all order method. We could not find any other calculations of $\alpha$ of the ground 
states of the Ge$^{2+}$ and Sn$^{2+}$ ions to make comparative analyses with our results.

\begin{table}[t]
\caption{\label{tab2} Dipole polarizabilities of the considered atomic systems are presented using different many-body
methods.}
\begin{ruledtabular}
\begin{tabular}{lccccc}
\textrm{System}& DF & MBPT(3) & RPA & LCCSD  & CCSD \\
\hline
B$^+$         &8.142   &9.720  & 11.374&11.875 & 10.413 \\
C$^{2+}$      &3.282   &3.804  & 4.503 &4.886   & 4.213 \\
Al$^+$        &19.514  &21.752 &26.289 &26.118  & 24.299 \\
Si$^{2+}$     &9.683   &10.482 &12.476 &12.847  & 11.893 \\
Zn            &37.317  &34.421 &50.846 &38.739  & 38.701 \\
Ga$^{+}$      &17.148  &15.796 &21.780 &19.138  & 18.455 \\
Ge$^{2+}$     &10.085  & 8.884 &12.011 &11.520  & 10.890 \\
Cd            &49.647  &35.728 &63.743 &45.086  & 45.898 \\
In$^{+}$      &25.734  &18.374 &29.570 &25.360  & 24.246 \\
Sn$^{2+}$     &16.445  &12.095 &17.941 &15.978  & 15.537 \\
\end{tabular}
\end{ruledtabular}  
\end{table}  

\begin{figure}[t]
 \includegraphics[width=5cm,angle=-90]{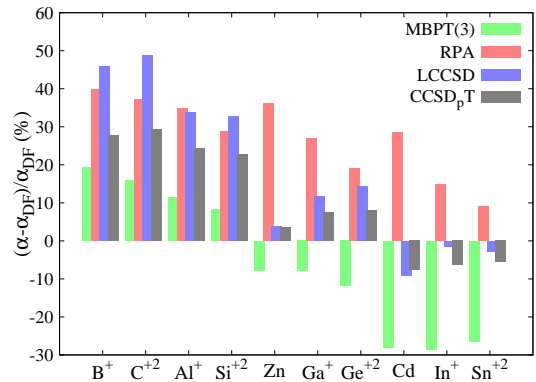}
 \caption{(color online) Histogram showing $(\alpha-\alpha_D)/\alpha_D$ (in \%) with different 
 many-body methods against the considered atomic systems.}
 \label{histo}
\end{figure}

\begin{figure}[t]
 \includegraphics[width=5cm,angle=-90]{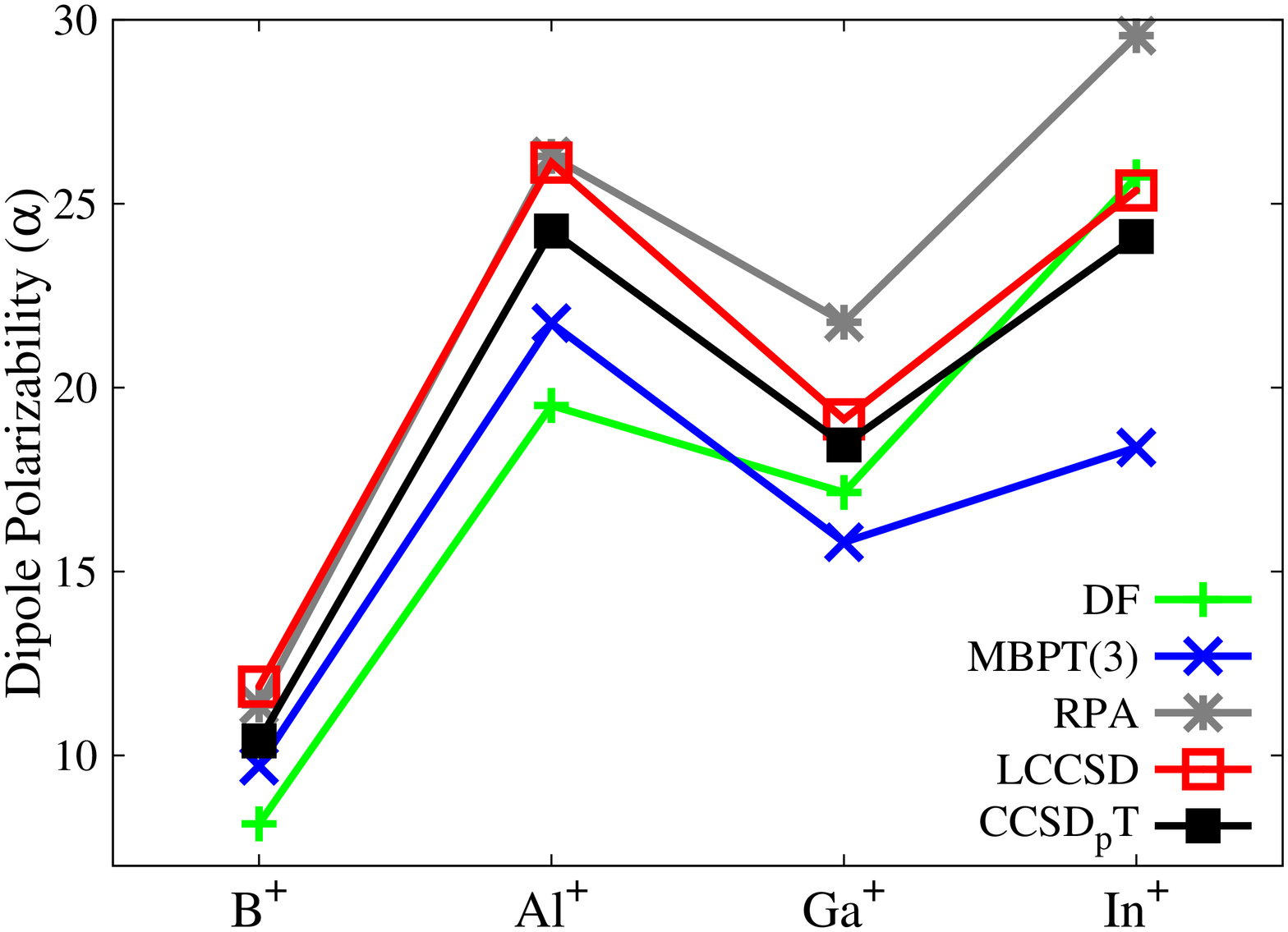}
  \caption{(color online) Trends in the calculations of dipole polarizabilities ($\alpha$) from the employed many-body methods
  in the considered singly charged ions.}
 \label{group1}
\end{figure}

\begin{figure}[t]
  \includegraphics[width=5cm,angle=-90]{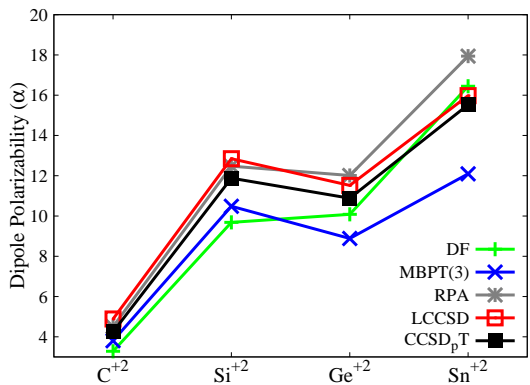}
  \caption{(color online) Trends in the calculations of dipole polarizabilities ($\alpha$) from the employed many-body methods
  in the considered doubly charged ions.}
 \label{group2}
\end{figure}

To assimilate the underlying roles of the electron correlation behavior in the evaluation of $\alpha$ 
of the ground states of the considered systems, we systematically present the calculated values of the 
dipole polarizabilities in Table \ref{tab2} from the DF, MBPT(3), RPA, LCCSD and CCSD methods. So, the 
differences between the CCSD results and the values quoted from the CCSD$_p$T method in Table \ref{tab1}
are the contributions from the partial triple excitations. Obviously, these differences are small in
magnitude implying that the contributions from the unaccounted higher order excitations are very small.
The lowest order DF results are smaller in magnitudes in the lighter systems but their
trends revert in the Cd isoelectronic systems with respect to their corresponding CCSD results.
Also, the MBPT(3) results do not follow a steady trend. In the B$^+$, C$^{2+}$, Al$^+$ and 
Si$^{2+}$ ions, the correlation effects enhance the $\alpha$ values in the MBPT(3) method from their
DF results while the MBPT(3) results are smaller than the DF values in the other systems. As has been
stated earlier RPA is a non-perturbative method embracing the core-polarization effects to all 
orders, but we find that the results are over estimated in this method compared to the CCSD results;
more precisely from the experimental values given in Table \ref{tab1}. We understand these differences
as the contributions from the pair-correlation effects that are absent in the RPA method, but they are
accounted intrinsically to all orders as the integral part of the CCSD method. The role of the 
pair-correlation effects in the determination of $\alpha$ are verified by examining contributions
from the individual MBPT(3) diagrams. The dominant contributing non-RPA diagrams appearing in the 
MBPT(3) method that take care of the pair-correlation effects are shown in Fig. \ref{mbpt}. In fact,
contributions from these non-RPA diagrams are found to be larger than the differences between the
RPA and CCSD results reported in Table \ref{tab2}. This finding advocates that there are large
cancellations among the lower order and higher order pair-correlation contributions in the CCSD method
bestowing modest size of contributions to $\alpha$, but they appear to be very significant in the
heavier systems to attribute accuracies in the results. To demonstrate the roles of the non-linear terms
to procure high precision $\alpha$ values in the considered ions, we have also given the results from
the LCCSD method in the above table. Although LCCSD is an all order perturbative method, but it clearly
omits higher order core-polarization and pair-correlation effects that crop-up through the non-linear 
terms involving $T^{(0)}T^{(0)}$ or higher powers of $T^{(0)}$. Consequently, this method also 
over estimates the results like the RPA method. The LCCSD results in B$^+$ and C$^{2+}$ are larger 
than the RPA values, but the LCCSD values are smaller than the RPA results in the other cases. This
clearly demonstrates intermittent trends of the correlation effects in the determination of $\alpha$
of the systems belonging to a particular group of elements in the periodic table to another through a given many-body
method as well as when they are studied using the methods with different levels of approximations. To 
manifest contributions from the correlations effects through various many-body methods quantitatively,
we portray the results obtained for $\alpha$ of the considered systems using these methods in a histogram 
as shown in Fig. \ref{histo}. This clearly bespeaks about the lopsided trend in the estimation of
$\alpha$ of the considered systems. Again, we also plot the $\alpha$ values of the singly and doubly 
charged ions separately in Figs. \ref{group1} and \ref{group2} in order to make a comparative analysis in the
propagation of correlation effects through the employed methods in these elements that belong to two different groups 
of the periodic table. This figure shows that the contributions from the correlation effects in the
singly charged and doubly charged ions do not exactly follow similar trends.

Finally, we would like to discuss about the trends in the correlation effects coming through various CCSD$_p$T 
terms. We give contributions explicitly from the individual CC terms of linear form and the rest as
``Others" in Table \ref{tab3}. Clearly, this table shows that the first term $DT^{(1)}$ gives the dominant  
contributions as it subsumes all the leading order core-polarization and pair-correlation effects
along with the DF result. The next dominant contributing term is $T_2^{(0)\dagger}DT_1^{(1)}$ which incorporates
some contributions from the correlation effects emanated at the MBPT(2) level and possess opposite
signs from the $DT^{(1)}$ contributions causing cancellations among them. It is also worthy to mention 
that contributions coming from the $T_2^{(0)\dagger}DT_2^{(1)}$ term corresponds to the higher order perturbation
and also accounts contributions from the doubly excited intermediate states. As seen from the table, 
these contributions are non-negligible suggesting that they should also be estimated accurately for
accomplishing high precision results and the sum-over-states approach may not be able to augment
these contributions suitably in the considered systems. Contributions from the other non-linear CC terms
at the final property evaluation level seem to be slender, although the differences between the LCCSD and
CCSD results emphasis their importance for accurate calculations of the atomic wave functions in the 
considered systems.

\begin{table}[t]
\caption{Contributions to the $\alpha$ of the ground state of considered atomic systems from various CCSD$_p$T terms.}
\begin{ruledtabular}
\begin{tabular}{lccccc}
\textrm{System}& \textrm{$DT_1^{(1)}$}& \textrm{$T_1^{(0)\dagger}DT_1^{(1)}$}& \textrm{$T_2^{(0)\dagger}
DT_1^{(1)}$}& \textrm{$T_2^{(0)\dagger}DT_2^{(1)}$}& \textrm{Others} \\
&+c.c&+c.c&+c.c&+c.c&\\
\hline                                                                                      \\
B$^+$         & 10.848 & $-$0.194  & $-$1.679 & 0.774  & 0.646  \\
C$^{2+}$      &  4.392 & $-$0.047  & $-$0.668 & 0.274  & 0.29   \\
Al$^+$        & 25.855 & $-$0.519  & $-$3.166 & 1.523  & 0.567  \\
Si$^{2+}$     & 12.589 & $-$0.160  & $-$1.475 & 0.666  & 0.260  \\
Zn            & 43.812 & $-$2.458  & $-$5.286 & 2.047  & 0.551  \\
Ga$^{+}$      & 20.223 & $-$0.545  & $-$2.409 & 0.837  & 0.335  \\
Ge$^{2+}$     & 11.846 & $-$0.198  & $-$1.363 & 0.476  & 0.122  \\
Cd            & 52.963 & $-$3.346  & $-$6.985 & 2.262  & 0.962  \\
In$^{+}$      & 27.134 & $-$0.882  & $-$3.647 & 1.064  & 0.441   \\
Sn$^{2+}$     & 17.249 & $-$0.366  & $-$2.286 & 0.603  & 0.326  \\
\end{tabular}
\end{ruledtabular}  
\label{tab3}
\end{table}  

\section{Conclusion}
We have employed a variety of many body methods to incorporate the correlation effects at 
different levels of approximations to unravel the role of the correlation effects and follow-up
their trends to achieve very accurate calculations of the dipole polarizabilities of three
groups of elements in the periodic table. We find the patterns in which the correlation effects behave
with respect to the mean-field level of calculations are divergent in the individual isoelectronic
systems through a particular employed many-body method. Also, our calculations reveal that inclusion of 
both the core-polarization and pair-correlation effects to all orders are equally important for
securing high precision dipole polarizabilities in the considered systems and the core-polarization 
effects play the pivotal role among them. Contributions from the doubly excited states are found
to be non-negligible implying that a sum-over-states approach may not be pertinent to carry out these studies. 
Our results obtained using the singles, doubles and important triples approximation in the coupled-cluster
method agree very well with the available experimental values in some of the systems except for
cadmium. In fact none of the reported theoretical results for cadmium agree with the measurement,
however there seem to be reasonable agreement among all theoretical results. This urges for further
experimental investigation of the cadmium result. In few systems, there are no experimental results
available yet and the reported precise values in the present work can be served as exemplars for
their prospective measurements.

\section{Acknowledgment}
BKS was supported partly by INSA-JSPS under project no. 
IA/INSA-JSPS Project/2013-2016/February 28, 2013/4098.
The computations reported in the present work were carried out 
using the 3TFLOP HPC cluster at Physical Research Laboratory, Ahmedabad.

\end{document}